\pdfoutput=1

\documentclass[superscriptaddress,amsmath,amssymb,twocolumn]{revtex4-1}

\usepackage{graphicx}
\usepackage{chemarr}
\usepackage{times,longtable}
\usepackage{hyperref}
\usepackage{color}
\usepackage{afterpage}
\usepackage{soul,comment}
\usepackage[usenames,dvipsnames]{xcolor}
\usepackage{colortbl}
\usepackage{pbox}
\usepackage{capt-of}

\graphicspath{{large/}}

\bibpunct{[}{]}{,}{n}{}{,}

\newcommand{\avg}[1]{\left\langle #1 \right\rangle}

\usepackage{chemarr}
\usepackage{epsf,mathtools}
\usepackage{graphicx}
\usepackage{dcolumn}
\usepackage{bm}

\begin{document}

\title{\textsf{Quantifying the entropic cost of cellular growth control}}

\author{Daniele De Martino}
\affiliation{Institute of Science and Technology Austria (IST Austria), Klosterneuburg (Austria)}

\author{Fabrizio Capuani}
\affiliation{Istituto Nazionale di Fisica Nucleare, Unit\`a di Roma 1, Rome (Italy)}

\author{Andrea De Martino}
\affiliation{Soft and Living Matter Lab, Institute of Nanotechnology (CNR-NANOTEC), Consiglio Nazionale delle Ricerche, Rome (Italy)}
\affiliation{Human Genetics Foundation, Turin (Italy)}

\begin{abstract}
We quantify the amount of regulation required to control growth in living cells by a Maximum Entropy approach to the space of  underlying metabolic states described by genome-scale models. Results obtained for {\it E. coli} and human cells are consistent with experiments and point to different regulatory strategies by which growth can be fostered or repressed. Moreover we explicitly connect the `inverse temperature' that controls MaxEnt distributions to the growth dynamics, showing that the initial  size of a colony may be crucial in determining how an exponentially growing population organizes the phenotypic space.
\end{abstract}

\maketitle

To a great extent, the physiologic state of a living cell is determined by how a large number of ``microscopic'' degrees of freedom subject to noise (nutrient import rates, metabolic reaction fluxes, gene expression levels, etc.) coordinate in response to the sensing of the extracellular conditions. This process ultimately correlates different regulatory variables and constrains the cell to a portion of the space of their  physico-chemically viable configurations. In turn, the amount by which the accessible volume of the feasible space is reduced (or the entropy change) can be thought to quantify, roughly speaking, the overall amount of regulation required to correctly modulate the cell's physiology. A key idea in such a scenario is that the desired fitness level and the strength of regulation necessary to achieve it are tightly linked \cite{bialek,book}. Having access to detailed information on genetic and metabolic variables one may now hope to characterize the set of phenotypes selected by regulation in terms of the behaviour of individual degrees of freedom. On the other hand, a principle-based approach might provide meaningful system-level insights. An important question in this respect is the following: is it possible to describe the selected region of the feasible space in physically precise terms? 

Bacteria, whose physiology is primarily described by their growth rate, may yield important clues. Experiments probing bacterial growth at single-cell resolution can in fact appraise the significant cell-to-cell variability that accompanies the establishment of a well-defined mean growth rate across an exponentially growing population \cite{jun,ken,elf}. Such a heterogeneity reflects the underlying phenotypes at a macroscopic level and may therefore carry strong regulatory signatures. Single-cell growth rate distributions measured for {\it E. coli} have indeed been shown to correspond to maximum entropy (MaxEnt) distributions of its viable metabolic flux patterns at fixed mean growth rate, suggesting that metabolic regulation realizes a tradeoff between the high fitness of fast-growing states and the high density of slow-growing ones \cite{physbio}. 

By shifting the optimization target from the growth rate to its entropic costs, the MaxEnt approach offers a view that is compatible both with the presence of noise in gene expression, which poses fundamental limits to growth rate optimization, and with the idea that the metabolic costs of strictly optimizing growth in fluctuating environments may be prohibitive. Here we show that the MaxEnt principle allows to quantify the strength of regulation required to both foster and repress growth from genome-scale models of metabolic networks, pointing to specific mechanisms that cells can exploit to implement those strategies. In addition, we propose an  interpretation for the `inverse temperature' parameter that controls MaxEnt distributions by connecting it to a well known population growth law.

We focus on metabolic degrees of freedom, denoting by $\mathbf{v}=\{v_i\}$ the fluxes of metabolic reactions ($i=1,\ldots,N$ with $N$ the number of reactions) and by $\lambda\equiv\lambda(\mathbf{v})$ the growth rate corresponding to flux configuration $\mathbf{v}$. The space $\mathcal{F}$ of feasible flux vectors $\mathbf{v}$ is formed by the non-equilibrium steady states of the underlying metabolic network, given by the solutions of $\mathbf{Sv=0}$, where $\mathbf{S}$ denotes the $M\times N$ stoichiometric matrix ($M$ being the number of chemical species) and a specific range of variability $[v_i^{\min},v_i^{\max}]$ is prescribed for each $v_i$ based on thermodynamic and kinetic constraints \cite{what}. MaxEnt distributions with prescribed mean growth rate $\avg{\lambda}$ over $\mathcal{F}$ are given by \cite{physbio}
\begin{equation} \label{dist}
p(\mathbf{v})=\frac{e^{\beta\lambda(\mathbf{v})}}{Z(\beta)}~~~~(\mathbf{v}\in\mathcal{F})~~,
\end{equation} 
where $Z(\beta)=\int e^{\beta\lambda(\mathbf{v})}d\mathbf{v}$ while $\beta$ is the Lagrange multiplier enforcing the constraint $\avg{\lambda}=\int\lambda(\mathbf{v})p(\mathbf{v})d\mathbf{v}$. The limit $\beta\to 0$ (resp. $\beta\gg 1$) describes a uniform distribution of flux vectors (resp. a distribution that concentrates around $\max_\mathbf{v}\lambda(\mathbf{v})\equiv\lambda_{\max})$. For each given $\beta$ (or $\avg{\lambda}$), sampling by (\ref{dist}) reduces the entropy of $\mathcal{F}$ with respect to the uniform sampling with $\beta=0$ by a factor $2^I$, with $I$ given by \cite{physbio}
\begin{equation}\label{entred}
I\log 2=\beta\avg{\lambda}-\int_0^\beta \avg{\lambda}d\beta'~~,
\end{equation}
where $\avg{\lambda}$ is a function of $\beta$. The above quantity (measured in bits) can be interpreted as the minimal  amount of regulation required to establish a given mean growth rate $\avg{\lambda}$. Clearly $I$ increases as cells are pushed to faster rates by increasing $\beta$. The $\avg{\lambda}$ vs $I$ curve obtained from (\ref{entred}) upon varying $\beta$ therefore separates the ($I,\avg{\lambda}$) plane in a feasible region (where $I$ is large enough for the corresponding value of $\avg{\lambda}$) and a forbidden region (where $I$ is too small). 

\begin{figure}[t!]
\begin{center}
\includegraphics[width=0.45\textwidth]{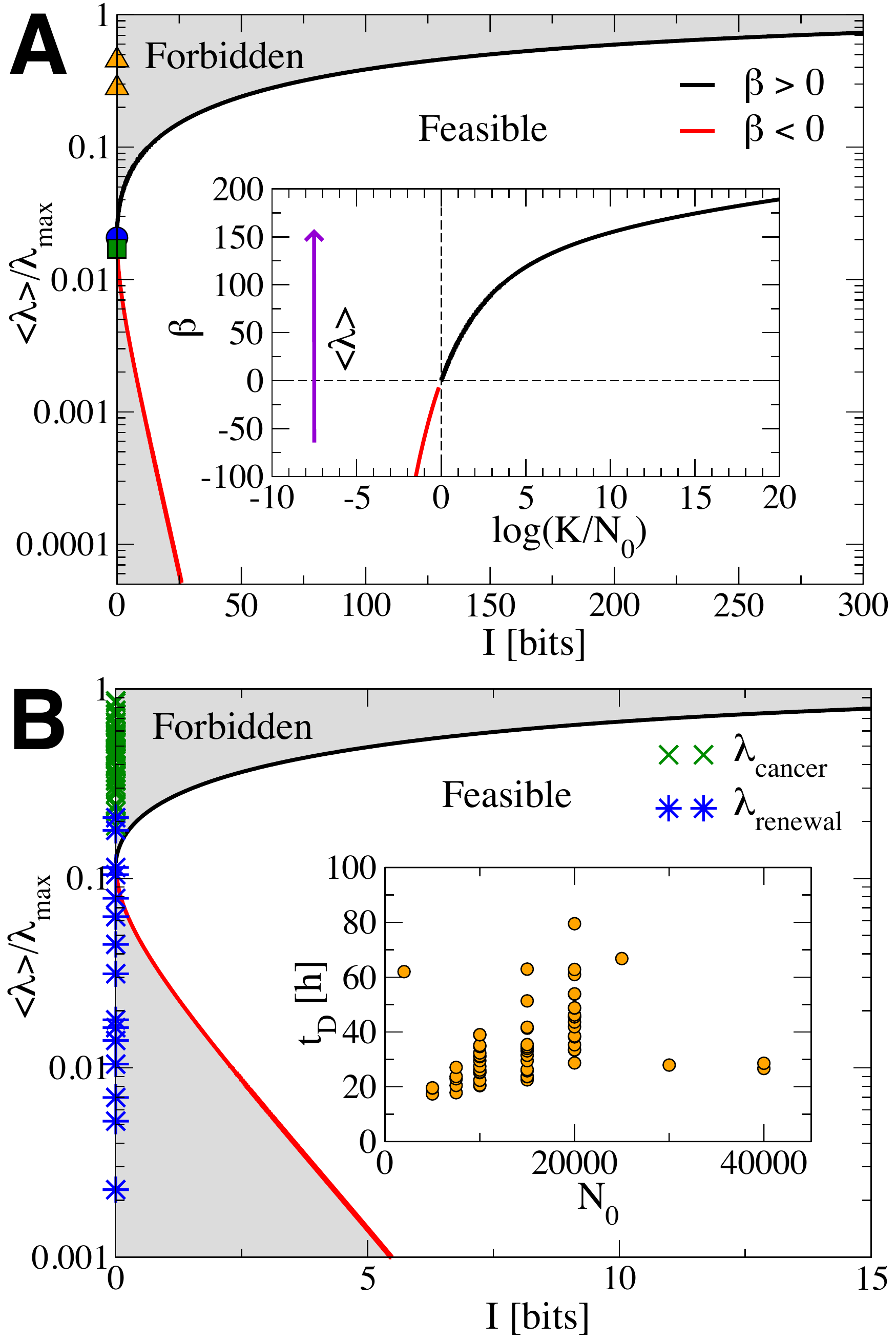}
\end{center}
\caption{\label{uno}Main panels: (A) $\avg{\lambda}/\lambda_{\max}$ versus $I$ trade-off curve computed from {\it E. coli}'s iJR904 genome-scale  metabolic network model assuming a glucose-limited minimal medium ($\lambda_{\max}=1/$h). The blue, green and orange markers denote respectively the values of $\avg{\lambda}/\lambda_{\max}$ (i) found for $\beta=0$ (i.e. for an unbiased sampling of the feasible space), (ii) estimated for {\it E. coli} in the human gut (roughly corresponding to a doubling time of 40 hours), and (iii) computed in \cite{physbio} for two sets of growth rate distributions, respectively described by the values $\avg{\lambda}/\lambda_{\max}\simeq 0.28$ and $\simeq 0.45$. (B) $\avg{\lambda}/\lambda_{\max}$ versus $I$ trade-off curve computed from the human catabolic core network with $\lambda_{\max}\simeq 0.046/{\rm h}$, corresponding to a fast growth rate for cancer cells (doubling time $\simeq 15$ hours). Blue and green markers represent respectively the estimated renewal rates of various healthy human tissues and the estimated growth rates of different types of cancers. Insets: (A) $\beta^\star$ versus $\log(K/N_0)$ for the minimal model of population growth described in the text (see Eq. (\ref{betastar})); (B) measured doubling times of different cancer cell types as functions of the inoculum size $N_0$. All data points were obtained from bionumbers.org except for those in the inset of panel (B), which were obtained from  \href{http://dtp.cancer.gov/discovery_development/nci-60/cell_list.htm}{dtp.cancer.gov/discovery\_development/nci-60/cell\_list.htm}.
}
\end{figure}

Figure \ref{uno}A (main panel, black line) shows the structure obtained by increasing $\beta$ starting from 0 (blue marker) for {\it E. coli} growth in a minimal glucose-limited medium based on the iJR904 genome-scale metabolic network model ($N=1075$, $M=761$ \cite{reed}).
Ref. \cite{physbio} has shown that MaxEnt distributions lying on this line reproduce different sets of empirical data by fitting a single parameter, with the values of $\avg{\lambda}/\lambda_{\max}$ displayed by orange markers. The standard {\it E. coli} growth rate in the human gut, corresponding to a doubling time of about 40 hours (green marker), instead appears to be close to the mean growth rate obtained for a flat sampling of $\mathcal{F}$ with $\beta=0$, with both lying close to 1\% of $\lambda_{\max}$. Such fitness values are achievable at very small regulatory costs, i.e. for $I\simeq 0$. This however implies that {\it slower} growth rates require some degree of regulation. A `slow growth' branch in the ($I,\avg{\lambda}$) diagram can  be obtained by simply computing (\ref{entred}) for $\beta<0$. This leads to the red curve in Fig. \ref{uno}A (main panel) and, in turn, to a second forbidden region at small $I$ and small $\avg{\lambda}$. 

It would be interesting to check whether cells explore the slow branch of the phase diagram as they seem to do with the fast branch. Perhaps unsurprisingly, though, few studies probed bacteria at very slow growth \cite{slowcoli}. In human cells, on the other hand, growth is actively downregulated. Fig. \ref{uno}B displays the $(\avg{\lambda},I)$ phase structure obtained from the carbon catabolic core metabolism of human cells \cite{capu} together with the estimated growth rates of 61 cancer types (green markers) and the estimated renewal rates of 21 human tissues (blue markers). The mean fitness obtained for $\beta=0$ is close to separating the two data sets, suggesting that healthy tissues might probe states close to the slow branch. 

\begin{figure}
\begin{center}
\includegraphics[width=0.48\textwidth]{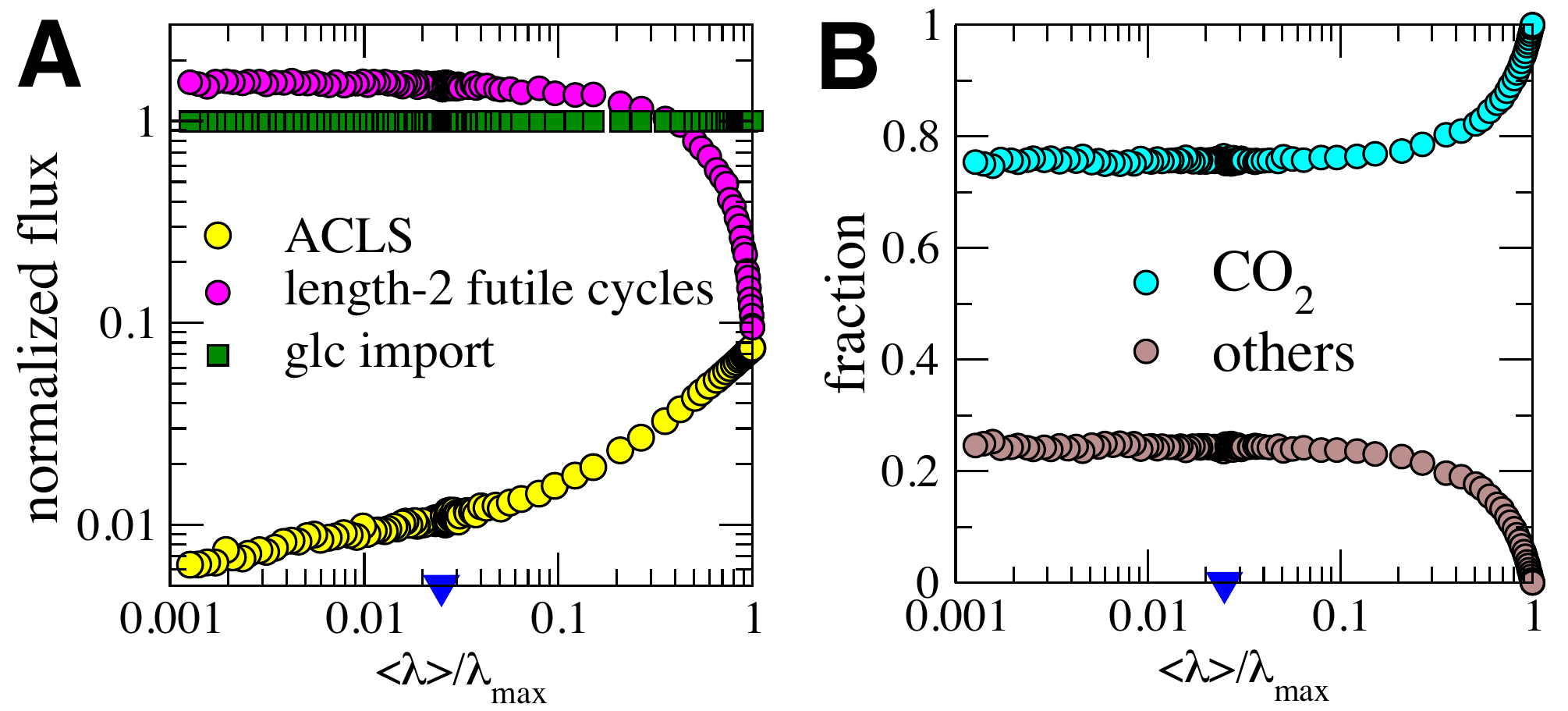}
\end{center}
\caption{\label{due}Modulation of different regulatory variables with $\avg{\lambda}/\lambda_{\max}$ for MaxEnt flux configurations of {\it E. coli} found under glucose-limited (maximum uptake $4.4$ mmol/g$_{\rm DW}$/h) aerobic conditions with $\lambda_{\max}\simeq 0.4/{\rm h}$. (A) Mean fluxes through acetolactate synthase (ACLS), length-2 futile cycles and glucose import. (B) Fraction of carbon excreted as CO$_2$ (cyan) and as other carbon-compounds (brown). All fluxes are normalized to the maximum glucose import flux. $\beta$ increases from $-\infty$ to $+\infty$ as $\avg{\lambda}/\lambda_{\max}$ grows. The blue triangle marks the value of $\avg{\lambda}/\lambda_{\max}$ corresponding to $\beta=0$.}
\end{figure}

It is instructive to compare $I$ to the dimension of $\mathcal{F}$, equal, in the case of {\it E. coli} presented in Fig. \ref{uno}A, to 233. 
In order to get close to $\lambda_{\max}$ cells have to invest considerably more than one bit per degree of freedom into regulation, in agreement with the view that growth rate maximization entails a finer and finer tuning of metabolic reactions (and higher regulatory costs). On the other hand, slowing growth below the ``unregulated'' limit only seems to require a fraction of a bit per degree of freedom. A careful look at solutions selected by the MaxEnt rule upon varying $\beta$ (which can be computed as in \cite{physbio}) sheds light on the regulatory pathways that cells modulate to adjust their fitness. Focusing on conditions for which $\lambda_{\max}\simeq 0.4/{\rm h}$ (so as to avoid effects due to gene expression costs that set in at faster rates \cite{cafba}), we see (Fig. \ref{due}A) that the flow through futile cycles anticorrelates with the mean growth rate for $\beta>0$ while it is roughly constant for $\beta<0$ (Fig. \ref{due}A), implying that the reduction  of chemical energy dissipation is a major mechanism of growth maximization \cite{physbio}. Likewise, increasing $\beta$ appears to select solutions for which CO$_2$ is the main carbon compound excreted (Fig. \ref{due}B), in agreement with the fact that the oxidative phenotype should be the dominating one at the growth rates under consideration. No major rearrangement of these pathways is observed for $\beta<0$, as the mean glucose intake also remains constant (Fig. \ref{due}A). This is not surprising as growth suppression requires much weaker regulation than growth optimization (Fig. \ref{uno}A). Interestingly, the only reaction that appears to be significantly modulated along the `slow branch' of the phase diagram is acetolactate synthase (ACLS, Fig. \ref{due}A), a key enzyme for the biosynthesis of branched-chain amino acids, which plausibly limits growth and is downregulated upon decreasing $\beta$. 

This analysis raises the question of whether the parameter $\beta$, which varies from $-\infty$ to $+\infty$, can be seen in more precise terms than simply as a `degree of optimization'. To get some insight, we consider a generalization of the logistic growth model defined in \cite{bara}. Consider a population of $N_0$ cells (indexed $i$) initially planted in a growth medium with finite carrying capacity $K$ and assume that their growth rates $\lambda_i$ are sampled independently from a distribution $q(\lambda)$ defined over the feasible space $\mathcal{F}$. If the number $n_i$ of cells with growth rate $\lambda_i$ evolves in time according to
\begin{equation}
\frac{\dot n_i}{n_i}=\lambda_i\left(1-\frac{N}{K}\right)~~~~~,~~~~~ N(t)=\sum_{i=1}^{N_0} n_i(t)~~,
\end{equation}
one easily sees that $n_i(t)=e^{\beta(t)\lambda_i}$ with
\begin{equation}
\beta(t)=t-\frac{1}{K}\int_0^t N(t')dt'~~.
\end{equation}
In turn, we can approximate $N(t)$ as
\begin{equation}
N(t)\simeq N_0 Z(\beta(t))~~~~~,~~~~~Z(\beta)=\int q(\lambda) e^{\beta \lambda}d\lambda~~,
\end{equation}
so that
\begin{equation}
\dot\beta=1-\frac{N_0}{K}Z(\beta)~~.
\end{equation}
At stationarity, $\beta$ settles to a value $\beta^\star$ fixed by the condition
\begin{equation}\label{betastar}
Z(\beta^\star)=\frac{K}{N_0}~~,
\end{equation}
which determines the asymptotic `degree of optimization' given $q(\lambda)$, the carrying capacity $K$ and the size of the inoculum $N_0$. The inset of Fig. \ref{uno}A shows the solutions obtained for $q(\lambda)\propto(1-\lambda/\lambda_{\max})^a$ \cite{physbio} as a function of $\log(K/N_0)$. One sees that $\beta>0$ for $K>N_0$ while it rapidly becomes more and more negative when $N_0>K$. In other terms, the asymptotic distribution is of the MaxEnt type, and concentrates on growth-suppressing states with $\beta<0$ for sufficiently `stressed' initial conditions.

Despite its crudeness, this setup hints at the drivers that cause a population of cells to organize in the feasible space with a specific value of $\beta$, providing in turn a dynamical justification of the MaxEnt scenario. In essence, it is the  exponential character of the growth law that leads to MaxEnt distributions. Interestingly, the size $N_0$ of the inoculum plays an important role in this process. Remarkably, the fact that $\langle\lambda \rangle$ decreases as $N_0$ increases is  consistent with the increase of doubling times with $N_0$ found for various cancers {\it in vitro} (at least for small enough $N_0$, see inset of Fig. \ref{uno}B). Overall, the MaxEnt approach that proved useful in other biological contexts \cite{mora} appears to provide a key to connect growth regulation to the entropy of the space of metabolic states. Our results also offer a simple physical interpretation of the parameter $\beta$. Experiments probing slow-growth regimes may clarify if {\it E. coli} cells saturate the slow branch of in Fig. \ref{uno}A as they seem to do with the fast branch.

~

\footnotesize{{\bf Acknowledgments} -- We are indebted with T. Gueudre for useful insights. The research leading to these results has received funding from the People Programme (Marie Curie Actions) of the European Union's Seventh Framework Programme (FP7/2007--2013) under REA grant agreement n. [291734].}


\end{document}